\documentclass[12pt]{article}
\usepackage{datetime,graphicx,amssymb,amstext,amsmath,parskip,abstract,titlesec}
\usepackage[top=30mm, bottom=30mm, left=20mm, right=20mm]{geometry}
\usepackage{setspace}
\numberwithin{equation}{section}

\allowdisplaybreaks

\begin{document}

\begin{titlepage}

\begin{center}

\vspace*{0.5cm}
\LARGE
\textbf{Momentum conserving defects in affine Toda field theories}

\vspace{1cm}

\large
\text{R. Bristow and P. Bowcock}

\emph{Department of Mathematical Sciences,\\
Durham University, Durham, U.K.\\
DH1 3LE}

\small
\text{\emph{Email:} rebecca.bristow@durham.ac.uk, peter.bowcock@durham.ac.uk}

\vspace{2cm}

\end{center}
\normalsize
A more general form of the defects with an extra degree of freedom (type II defects) introduced previously is investigated. Conditions on the form of the defect are found which ensure that a system containing a defect has a momentum-like conserved quantity. The defect equations of motion plus some easily found extra equations, when taken to hold everywhere, give a B\"{a}cklund transformation between the bulk theories on either side of the defect. This strongly suffests that such systems are integrable. Momentum conserving defects in affine Toda field theories based on the $A_n$, $B_n$, $C_n$ and $D_n$ series of Lie algebras are found. The delays of solitons passing through a defect in the $D_4$ affine Toda field theory are calculated.
\end{titlepage}

\section{Introduction} \label{sec:intro}

It was found in \cite{bcz04a,bcz04b} that it is possible for some two-dimensional integrable field theories to accommodate discontinuities in the fields and yet remain integrable. This discontinuity is referred to as a defect in the theory, and the fields on either side of the discontinuity are related by some set of defect conditions. There may be a potential and extra degrees of freedom which exist only at the defect and influence the defect conditions. Here we consider a Lagrangian set-up, where the Lagrangian density contains a term for the bulk theory on either side of the defect (confined to the appropriate region) and a defect term which is confined to a single point.

Affine Toda field theories (ATFTs) were first introduced in \cite{tod70} when a one-dimensional chain of particles with nearest neighbour interactions was investigated. The potential of this system could be written in such a way that it depended upon the simple roots of affine $A_n$. This was later modified to be dependent on both $x$ and $t$ \cite{mik79}, and then generalised to give field theories based on the roots of any affine Lie algebra \cite{bog76,mop81}. Such theories are integrable \cite{mik79,mop81,ot85} and soliton solutions have been found \cite{hol92,mm93,mcg94b}.

Some of the earliest studies of defects were in quantum integrable field theories, for example in a free fermion theory \cite{dms94a,dms94b} and in sine-Gordon theory \cite{kl99}, and here it was shown that integrable defects must be purely reflecting or transmitting. Classical purely transmitting defects first appeared in \cite{bcz04a}, where the Lagrangian approach to classical defects used in this paper was pioneered. For the defects investigated in \cite{bcz04a,bcz04b,cz09a} the bulk fields couple to each other at the defect. These are type I defects and allow discontinuities in the fields at the defect, provided that the defect conditions are satisfied. Initially momentum and energy conservation, rather than full integrability with an infinite number of conserved charges in involution, were all that was investigated \cite{bcz04a}. Despite the defect breaking translational invariance it was found that, for particular type I defects, it is possible for such systems to have conserved momentum. However these momentum conserving defects are only compatible with an $A_n$ ATFT in the bulk \cite{bcz04b}. Constructing the Lax pair showed that the restrictions on the defect which ensured energy and momentum conservation were necessary and sufficient to ensure the existence of an infinite number of conserved charges \cite{bcz04b,cz09a}. The sine-Gordon and $A_2$ type I defects have been shown to be integrable \cite{hk08,doi15}.

Since investigations of solitons and the integrability of ATFTs in the bulk found that the results for ATFTs based on different sets of simple roots were closely linked it would not be unreasonable to expect the type I defects to provide integrable defects for all ATFTs. Unfortunately this is not the case, as it seems only ATFTs based on $A_n$ can support a type I defect whilst still remaining a momentum conserving (and so likely integrable) system \cite{bcz04b}. However, in \cite{cz09b} a modification was proposed which allowed a momentum conserving defect to appear within the Tzitz\'{e}ica model (excluded from the integrable type I defects due to being based on folded $A_2$ roots rather than purely on $A_n$). This defect, referred to as a type II defect, introduced an additional degree of freedom only at the defect. Although this type II defect has not been explicitly shown to be integrable there is a strong body of evidence to suggest that it is, namely that momentum and energy are conserved and requiring momentum conservation gave sufficient constraints on the defect to ensure it was integrable in the sine-Gordon and $A_2$ cases, solitons were able to pass through it with no change other than a delay (determined by the rapidity of the soliton and the defect parameters), and that the existence of an infinite number of conserved charges has been shown for the Tzitz\'{e}ica defect \cite{aagz11}. Liouville integrability of defects with additional degrees of freedom has been investigated in \cite{ad12b}.

We will attempt to generalise these type II defects to accommodate any number of bulk fields and degrees of freedom at the defect in the hope of finding momentum conserving defects for all ATFTs. Since energy-momentum conservation was found to be such a powerful tool in the type I case we will be looking for momentum conserving defects rather than integrable defects, as it appears likely that they will be the same thing.

Such defects are also of interest because of their link with B\"{a}cklund transformations. In \cite{bcz04a, bcz04b} it was noted that the defect conditions of any momentum conserving type I defect in an $A_n$ ATFT were a B\"{a}cklund transformation if the defect conditions were taken to hold everywhere. In \cite{cz09b} a new B\"{a}cklund transformation for the Tzitz\'{e}ica model was found from the defect conditions. In this paper we show that the defect conditions of a momentum conserving defect can always be augmented to provide a set of equations which are a B\"{a}cklund transformation for the bulk theory. If the defect equations linking the theories on either side are a B\"{a}cklund transformation then we would expect the system to have soliton solutions which pass through the defect, a feature of integrable systems.

\section{Momentum conservation and the generalised type II defect} \label{sec:momconsv}

In this section we shall derive conditions for a general class of type II defects to be momentum conserving.
Type II defects were introduced in \cite{cz09b}. They differed from type 1 defects in having a single extra degree of freedom confined to the defect in addition to the fields in the bulk. This extra freedom allowed the authors to construct a momentum conserving defect for the Tzitz\'{e}ica model, something which had not been possible within the framework of type 1 defects. Here we shall generalise the results in \cite{cz09b} by considering the situation where there are any number of bulk fields and any number of extra degrees of freedom confined to the defect.

In what follows we take the defect to lie at $x=0$. The bulk vector fields in the region $x\leq 0$ will be called $u(x,t)$, the bulk vector fields  in the region $x\geq 0$ will be called $v(x,t)$ and the degrees of freedom living on the defect at $x=0$ are labelled $\lambda(t)$. We shall refer to the $\lambda(t)$ as {\it auxiliary fields}. (The term field may seem a peculiar choice as $\lambda$ has no spatial dependence; however when we come to consider B\"{a}cklund transformations in the next section we will see that it is natural to extend the definition of $\lambda$ to take values in the bulk.)  Each of $u$, $v$ and $\lambda$ is a vector field and we denote their components as $u_1,u_2,...$, $v_1,v_2,...$, $\lambda_1,\lambda_2,...$. Additionally we will assume that $u$ and $v$ describe two copies of the same bulk theory but on different sides of the defect, so that the number of components of $u$ and $v$ are equal. There may be any number of components of the auxiliary vector field $\lambda$.

The Lagrangian description of the theory in the presence of a defect at $x=0$ is given in terms of a density
\begin{align}
\mathcal{L} = \Theta(-x)\mathcal{L}^{u} +\Theta(x)\mathcal{L}^{v} +\delta(x)\mathcal{L}^D, \label{eq:Lgen}
\end{align}
where the bulk Lagrangian densities
\begin{align}
\mathcal{L}^{u}&=\frac{1}{2}(u_{i,t}u_{i,t}-u_{i,x}u_{i,x})-U(u)\label{eq:bulkLu} \\
\mathcal{L}^{v}&=\frac{1}{2}(v_{i,t}v_{i,t}-v_{i,x}v_{i,x})-V(v). \label{eq:bulkLv}
\end{align}
govern the behaviour of the bulk fields $u$ and $v$. Subscripts of $t$ and $x$ denote partial differentiation with respect to that variable and are separated from subscripts of indices by a comma. Einstein sum notation is used throughout. The two bulk theories are coupled at $x=0$ via the  defect Lagrangian $\mathcal{L}^D$ which depends on $u$, $v$ and $\lambda$. 

The form of $\mathcal{L}^D$ we will consider in the present work is motivated by combining features from existing examples of defects. An example of a type I defect coupling multicomponent fields $u$ and $v$ is the defect for $A_n$ ATFT considered in \cite{cz09a}; its Lagrangian is of the form
\begin{align}
\mathcal{L}^D =& \frac{1}{2}u_iA_{ij}u_{j,t} +\frac{1}{2}v_iA_{ij}v_{j,t} +u_i(I-A)_{ij}v_{j,t} -F(u,v) \label{eq:type1}
,\end{align}
where $A$ is a constant, antisymmetric matrix. The type II defect for the Tzitz\'{e}ica model considered in \cite{cz09b} is of the form 
\begin{align}
\mathcal{L}^D =& uv_t +2\lambda\left(u_t-v_t\right) -F(u,v,\lambda) \label{eq:type2}
,\end{align}
where $u$, $v$ and $\lambda$ are scalar fields.

In both of these examples, the defect Lagrangian consists of two parts: a defect potential $F=F(u,v,\lambda)$ and `kinetic terms' coupling the time derivatives of the fields to the fields themselves via constant matrices. In this paper we shall consider the most general defect of this form, combining the vector field aspect of the type I defect (which allowed it to encompass all $A_n$ ATFTs) with the auxiliary field appearing in the type II defect (which allowed a momentum conserving defect to be constructed for an ATFT not based on $A_n$). The work in \cite{rob14a} went some way toward combining the two approaches, but required the number of auxiliary fields to be equal to or a multiple of the number of bulk fields. The defect Lagrangian density we consider is
\begin{align}
\mathcal{L}^D =& \frac{1}{2}u_iA_{ij}u_{j,t} +\frac{1}{2}v_iB_{ij}v_{j,t} +u_iC_{ij}v_{j,t} \nonumber \\
&+\frac{1}{2}\lambda_iW_{ij}\lambda_{j,t} +\lambda_iX_{ij}u_{j,t} +\lambda_iY_{ij}v_{j,t} -F(u,v,\lambda) \label{eq:L}
,\end{align}
where $A$, $B$, $C$, $W$, $X$ and $Y$ are arbitrary, constant, real coupling matrices. 
This general form of defect Lagrangian depends on a plethora of unknown couplings contained in the matrices $A$, $B$, $C$, $W$, $X$ and $Y$. The main purpose of this section will to be use the freedom to make field redefinitions and the constraints arising from the condition that the defect conserve momentum to pin down the form of this Lagrangian much more precisely.

We can immediately see that some of the couplings in the defect Lagrangian \eqref{eq:L} are redundant. The matrices $A$, $B$ and $W$ can be taken to be antisymmetric as any symmetric part simply adds a total derivative to the Lagrangian which is physically irrelevant, at least in the classical case. Further simplifications can be made by using field redefinitions to put the Lagrangian in a canonical form. The form of the Lagrangian is not altered under the redefinition of the auxiliary fields $\lambda_i \rightarrow \alpha_{ij}u_j +\beta_{ij}v_j +\gamma_{ij}\lambda_j$, where $\alpha$ and $\beta$ are any matrices and $\gamma$ is an invertible matrix to ensure the degrees of freedom associated to the auxiliary fields are not removed, because $\lambda$ does not appear in the bulk Lagrangian. The bulk fields can also be transformed as $u_i \rightarrow Q_{ij}u_j$, $v_i \rightarrow Q'_{ij}v_j$ without changing the general form of the bulk and defect Lagrangians provided $Q$ and $Q'$ are orthogonal. We intend to use these field redefinitions to simplify the Lagrangian in eq.\eqref{eq:L} as far as possible, `absorbing' the freedom in the arbitrary coupling matrices into the auxiliary fields. We will find that any momentum conserving defect of the form given above is equivalent, up to some field redefinitions, to a defect in which each component of the fields may couple in either the type I or the type II manner seen in eqs.\eqref{eq:type1},\eqref{eq:type2}.

We begin by further simplifying $W$, the antisymmetric matrix containing the couplings between auxiliary fields. The spectral theorem states there exists a change of basis  $\lambda_i \rightarrow \gamma_{ij}\lambda_j$ where the matrix $\gamma$ is orthogonal, in which the antisymmetric matrix $W$ takes the block-diagonal form
\begin{align}
W\to \gamma^T W \gamma=
\left(\begin{matrix}
0 & l_1 & \hdots & 0 & 0 & 0 & \hdots & 0 \\
-l_1 & 0 & \hdots & 0 & 0 & 0 & \hdots & 0 \\
\vdots & \vdots & \ddots & \vdots & \vdots & \vdots & & \vdots \\
0 & 0 & \hdots & 0 & l_k & 0 & \hdots & 0 \\
0 & 0 & \hdots & -l_k & 0 & 0 & \hdots & 0 \\
0 & 0 & \hdots & 0 & 0 & 0 & \hdots & 0 \\
\vdots & \vdots & & \vdots & \vdots & \vdots & \ddots & \vdots \\
0 & 0 & \hdots & 0 & 0 & 0 & \hdots & 0 \\
\end{matrix}\right)
.\end{align}
where the matrix has $2k$ non-zero eigenvalues, $\pm i l_j$. We can also scale the auxiliary fields $\lambda_i\rightarrow c_i\lambda_i$, where $c_i$ are some scalars, to take all entries in this block-diagonal matrix to $\pm 1$. These field redefinitions can be carried out without loss of generality, and so we can always use them to set
\begin{align}
W=\left(\begin{matrix}
0 & 1 & \hdots & 0 & 0 & 0 & \hdots & 0 \\
-1 & 0 & \hdots & 0 & 0 & 0 & \hdots & 0 \\
\vdots & \vdots & \ddots & \vdots & \vdots & \vdots & & \vdots \\
0 & 0 & \hdots & 0 & 1 & 0 & \hdots & 0 \\
0 & 0 & \hdots & -1 & 0 & 0 & \hdots & 0 \\
0 & 0 & \hdots & 0 & 0 & 0 & \hdots & 0 \\
\vdots & \vdots & & \vdots & \vdots & \vdots & \ddots & \vdots \\
0 & 0 & \hdots & 0 & 0 & 0 & \hdots & 0 \\
\end{matrix}\right)
.\end{align}
The field redefinition on $\lambda$ will also affect the matrices $X$ and $Y$ but these can be ignored as they amount to redefinitions of what are already arbitrary matrices. With $W$ as above, the components of the  auxiliary field $\lambda_i$ naturally divide into those for $i=1\dots 2k$ which couple to other auxiliary fields, and the remaining components in the zero eigenspace of $W$ which have no coupling to other auxiliary fields in the `kinetic' part of the defect Lagrangian. The components of $\lambda$ which couple to other auxiliary fields are relabelled as $\xi_1,\xi_2,...$, components of the field vector $\xi$, and the components of $\lambda$ which couple to no other auxiliary fields are relabelled as $\mu_1,\mu_2,...$, components of the field vector $\mu$. In terms of $\xi$ and $\mu$ the defect Lagrangian density can now be rewritten as
\begin{align}
\mathcal{L}^D =& \frac{1}{2}u_iA_{ij}u_{j,t} +\frac{1}{2}v_iB_{ij}v_{j,t} +u_iC_{ij}v_{j,t} +\frac{1}{2}\xi_{i}W_{ij}\xi_{j,t} \nonumber \\
&+\mu_{i}X_{ij}u_{j,t} +\xi_{i}\hat{X}_{ij}u_{j,t} +\mu_{i}Y_{ij}v_{j,t} +\xi_{i}\hat{Y}_{ij}v_{j,t} -F \label{eq:Lredef1}
\end{align}
where matrices $X$ and $Y$ have been split into the smaller matrices $X$, $\hat{X}$, $Y$ and $\hat{Y}$ in order to separate the couplings of the bulk fields to $\{\mu_i\}$ and $\{\xi_i\}$, and the matrix $W$ is from now on taken to be
\begin{align}
W=\left(\begin{matrix}
0 & 1 & \hdots & 0 & 0 \\
-1 & 0 & \hdots & 0 & 0 \\
\vdots & \vdots & \ddots & \vdots & \vdots \\
0 & 0 & \hdots & 0 & 1 \\
0 & 0 & \hdots & -1 & 0 \\
\end{matrix}\right). \label{eq:W}
\end{align}

Having simplified $W$ as far as we can we now turn to the couplings of $\xi$ to the bulk fields. The redefinitions $\xi_{i} \rightarrow W_{ij}\hat{X}_{jk}u_k +W_{ij}\hat{Y}_{jk}v_k +\xi_{i}$ give
\begin{align}
\frac{1}{2}\xi_{i}W_{ij}\xi_{j,t}\rightarrow\frac{1}{2}\left(W_{ik}\hat{X}_{kl}u_l +W_{ik}\hat{Y}_{kl}v_l +\xi_{i}\right)W_{ij}\left(W_{jk}\hat{X}_{kl}u_{l,t} +W_{jk}\hat{Y}_{kl}v_{l,t} +\xi_{j,t}\right)
.\end{align}
Using $W^2=-I$ it is then straightforward to show that this provides cancellations which leave the Lagrangian density as
\begin{align}
\mathcal{L}^D =& \frac{1}{2}u_iA_{ij}u_{j,t} +\frac{1}{2}v_iB_{ij}v_{j,t} +u_iC_{ij}v_{j,t} +\frac{1}{2}\xi_{i}W_{ij}\xi_{j,t} +\mu_{i}X_{ij}u_{j,t} +\mu_{i}Y_{ij}v_{j,t} -F \label{eq:Lredef2}
.\end{align}
As before the effect of these field redefinitions on the arbitrary matrices $A$, $B$ and $C$ has been negated by an appropriate redefinition of these matrices.

This is the canonical form for the Lagrangian that we shall work with henceforth. We shall now look for the conditions 
on the matrices $A$, $B$, $C$, $W$, $X$ and $Y$ and potential $F$ which arise from demanding that the system described by the Lagrangian in eq.\eqref{eq:Lredef2} has a conserved momentum and energy. We expect that demanding momentum conservation will be sufficient to ensure the integrability of the defect. 

The Euler-Lagrange equations arising from the Lagrangian density in eq.\eqref{eq:Lgen} with the defect Lagrangian in eq.\eqref{eq:Lredef2} give the equations of motion
\begin{align}
& & x&<0: & 0 =& u_{i,tt}-u_{i,xx}+U_{u_i} & & \label{eq:eombulk1} \\
& & x&>0: & 0 =& v_{i,tt}-v_{i,xx}+V_{v_i} & & \label{eq:eombulk2} \\
& & x&=0: &u_{i,x} =& A_{ij}u_{j,t} +C_{ij}v_{j,t} -X_{ji}\mu_{j,t} -F_{u_i} \label{eq:eomdefect4} \\
& & & &v_{i,x} =& C_{ji}u_{j,t} -B_{ij}v_{j,t} +Y_{ji}\mu_{j,t} +F_{v_i} \label{eq:eomdefect5} \\
& & & &0 =& W_{ij}\xi_{j,t} -F_{\xi_{i}} \label{eq:eomdefect6} \\
& & & &0 =& X_{ij}u_{j,t} +Y_{ij}v_{j,t} -F_{\mu_{i}} \label{eq:eomdefect7}
,\end{align}
where a subscript containing a field denotes partial differentiation with respect to that field.

The total energy of the fields in the bulk is
\begin{align}
E =& \int_{-\infty}^0 \!\!\!\! dx \left( \frac{1}{2}\left( u_{i,t}u_{i,t} +u_{i,x}u_{i,x} \right) +U \right) +\int_0^{\infty} \!\!\!\! dx \left( \frac{1}{2}\left( v_{i,t}v_{i,t} +v_{i,x}v_{i,x} \right) +V \right) \label{eq:E}
\end{align}
and we expect the conserved total energy to be the sum of this bulk energy plus some contribution from the defect. Differentiating eq.\eqref{eq:E} with respect to $t$ and then using the bulk equations of motion in eqs.\eqref{eq:eombulk1}, \eqref{eq:eombulk2} to rewrite the integrand as a total $x$ derivative allows us to carry out the integration (with $\{u_i\},\{v_i\}\rightarrow \text{constant}$ and $U,V \rightarrow 0$ as $x\rightarrow \pm\infty$), giving
\begin{align}
\frac{\mathrm{d}E}{\mathrm{d}t} =& \left.\left(u_{i,x}u_{i,t} -v_{i,x}v_{i,t}\right)\right|_{x=0} \label{eq:Et}
.\end{align}
In order for this term to be conserved we must be able to write the right hand side of this equation as a total time derivative. Using the defect conditions in eqs.\eqref{eq:eomdefect4}, \eqref{eq:eomdefect5} to remove the $x$ derivatives we find that eq.\eqref{eq:Et} may be rewritten as
\begin{align}
\frac{\mathrm{d}E}{\mathrm{d}t} = -\frac{\mathrm{d}F}{\mathrm{d}t} \label{eq:EtFt}
.\end{align}
Therefore $E+F$ is the conserved energy-like quantity, where $E$ is the bulk energy and $F$ is the defect potential. The introduction of a defect at $x=0$ does not break the time translation symmetry of the system, so perhaps it is not surprising that it is always possible to construct a conserved energy without placing any further constraints on the couplings in the defect Lagrangian.

Since the defect breaks manifest translation invariance, the system is no longer obviously momentum conserving, and we expect requiring conservation of momentum to be far more restrictive than requiring conservation of energy. Total momentum of the fields in the bulk is given by
\begin{align}
P =& \int_{-\infty}^0 \!\!\!\! dx \left( u_{i,x}u_{i,t} \right) +\int_0^{\infty} \!\!\!\! dx \left( v_{i,x}v_{i,t} \right) \label{eq:P}
\end{align}
and again we will require that this plus some defect contribution is conserved.  Differentiating eq.\eqref{eq:P} with respect to $t$, using the bulk equations of motion in eqs.\eqref{eq:eombulk1}, \eqref{eq:eombulk2} to rewrite the integrand as a total $x$ derivative and carrying out the integration gives
\begin{align}
\frac{\mathrm{d}P}{\mathrm{d}t} = \left.\left(\frac{1}{2}\left( u_{i,t}u_{i,t} +u_{i,x}u_{i,x} -v_{i,t}v_{i,t} -v_{i,x}v_{i,x} \right) -U +V \right)\right|_{x=0} \label{eq:Pt}
.\end{align}
In order for the system to be momentum conserving we must be able to rewrite eq.\eqref{eq:Pt} as
\begin{align}
\frac{\mathrm{d}P}{\mathrm{d}t}=-\frac{\mathrm{d}\Omega}{\mathrm{d}t} \label{eq:PtOt}
\end{align}
where $\Omega$ is the defect contribution to the total momentum of the system. Using the defect conditions in eqs.\eqref{eq:eomdefect4}-\eqref{eq:eomdefect7} we now aim to find the restrictions on the couplings at the defect and the defect potential which are necessary to ensure the system is momentum conserving and so (hopefully) integrable. In order for eq.\eqref{eq:Pt} to be written as a total time derivative the $x$ derivatives must be removed, which can only be done by substituting in eqs.\eqref{eq:eomdefect4},\eqref{eq:eomdefect5}. This gives
\begin{align}
\frac{\mathrm{d}P}{\mathrm{d}t} =&
\frac{1}{2}u_{i,t}\left(I-A^2-CC^T\right)_{ij}u_{j,t}
-\frac{1}{2}v_{i,t}\left(I-B^2-C^TC\right)_{ij}v_{j,t}
-u_{i,t}\left(AC-CB\right)_{ij}v_{j,t}
\nonumber \\
& +u_{i,t}\left(AX^T-CY^T\right)_{ij}\mu_{j,t}
-v_{i,t}\left(C^TX^T+BY^T\right)_{ij}\mu_{j,t}
+\frac{1}{2}\mu_{i,t}\left(XX^T-YY^T\right)_{ij}\mu_{j,t} \nonumber \\
& -\left(F_{u_i}A_{ij}+F_{v_i}C^T_{ij}\right)u_{j,t} -\left(F_{u_i}C_{ij}-F_{v_i}B_{ij}\right)v_{j,t}
+\left(F_{u_i}X^T_{ij}-F_{v_i}Y^T_{ij}\right)\mu_{j,t} \nonumber \\
& +\frac{1}{2}\left(F_{u_i}F_{u_i}-F_{v_i}F_{v_i}\right) -U +V \nonumber \\
&+\left (-\xi_{k,t}W_{ki} -F_{\xi_{i}}\right)\left(\rho_i+\tau_{ij}u_{j,t}+\phi_{ij}v_{j,t}\right)\nonumber \\
&+\left (u_{k,t}X^T_{ki} +v_{k,t}Y^T_{ki} -F_{\mu_{i}}\right)\left(\sigma_i+\pi_{ij}u_{j,t}+\chi_{ij}v_{j,t}+\psi_{ij}\mu_{j,t}+\omega_{ij}\xi_{j,t}\right )
\label{eq:Pt2} 
.\end{align}
For the right hand side of this equation to be a total time derivative we must remove all terms which are not linear in time
derivatives of the fields. In the last two lines of this equation we have used the freedom to add multiples of the expressions in eqs.\eqref{eq:eomdefect6}-\eqref{eq:eomdefect7} which vanish as a consequence of the equations of motion. We have not added multiples of the expressions in eqs.\eqref{eq:eomdefect4}-\eqref{eq:eomdefect5}, as these would reintroduce 
derivatives of the fields with respect to $x$ which cannot be expressed as time derivatives. Equally the multiplicative 
factors of the expressions in eqs.\eqref{eq:eomdefect4}-\eqref{eq:eomdefect5} have been chosen to introduce no higher
than quadratic terms of time derivatives of fields into eq.\eqref{eq:Pt2} as these also could not be made into a total time 
derivative. They also must not introduce any quadratic terms which do not appear elsewhere in the expression, as such terms would have nothing to cancel with, cannot be written as a total time derivative, and so would immediately be set to zero.

Let us begin by considering the term $\mu_{i,t}\left(XX^T-YY^T\right)_{ij}\mu_{j,t}$. For this 
to be a  total time derivative it must identically vanish, and as the quantity $XX^T-YY^T$ is explicitly symmetric, we have that  $XX^T=YY^T$. Now consider the case in which a particular auxiliary field decouples from $u$ but not from $v$. It is always possible to permute the labels on the fields $\{\mu_{i}\}$ by a field redefinition so that the field $\mu_1$ is the one decoupling from $u$ but not from $v$, so $X_{1j}=0 \:\: \forall \: j$. The condition $X_{ij}X^T_{jk}=Y_{ij}Y^T_{jk}$ then requires $Y_{1j}Y^T_{jk}=0 \:\: \forall \: k$. One of the conditions from this is $Y_{1j}Y_{1j}=0$ and since all coupling matrices are assumed to be real this is only satisfied if $Y_{1j}=0 \:\: \forall \: j$. Therefore if an auxiliary field decouples from $u$ it must also decouple from $v$ and vice versa. From eq.\eqref{eq:eomdefect7} we then have that the equation of motion of the field $\mu_1$ is then $F_{\mu_1}=0$, and so if an auxiliary field decouples completely from all other auxiliary fields and from one of the bulk field vectors then it disappears from the defect Lagrangian.

Now consider the $\mu_{i,t}X_{ij}u_{j,t}+\mu_{i,t}Y_{ij}v_{j,t}$ terms. We take vectors $u$ and $v$ to have $n$ components and vector $\mu$ to have $m$ components. The matrix $X^T$ has a kernel which will be some subspace of the vector space $\mu$ is living in. By a transformation of $\mu$ we can take the basis of the kernel of $X^T$ to be the final $k$ elements of $\mu$. After this transformation the final $k$ columns of $X^T$ will be zero. The final $k$ components of $\mu$ completely decouple from $u$, and so by the argument in the above paragraph they also completely decouple from $v$, and so $Y^T$ also has the final $k$ columns as zero. The final $k$ components of $\mu$ are now auxiliary fields which completely decouple from $u$ and $v$, and so can be removed from the Lagrangian. The vector $\mu$ is now length $m-k$ and the matrices $X^T$ and $Y^T$ must have a kernel of $0$, otherwise further $\mu$ components should have decoupled. A matrix can only have a zero kernel if the number of rows is greater than or equal to the number of columns. So $X$ and $Y$ are both $(m-k)\times n$ matrices with $m-k\leq n$. The matrix $X$ also has a kernel, and we can take this to have a basis consisting of the first $r$ components of $u$ by an orthogonal transformation of $u$. These components of $u$ completely decouple from the auxiliary fields, and so we choose to denote the vector containing only these components of $u$ as $u^{(1)}$, where the superscript indicates that these fields couple like a type I defect. We will call the vector containing the remaining components of $u$ $u^{(2)}$. The first $r$ columns of $X$ are then zero, and by rewriting the term $\mu_{i,t}X_{ij}u_{j,t}$ as $\mu_{i,t}\left( 0 \:  X \right)_{ij}u_{j,t}=\mu_{i,t}X_{ij}u^{(2)}_{j,t}$ we have that $X$ is a $(m-k)\times(n-r)$ matrix with zero kernel and so $n-r\leq m-k$. But if $n-r<m-k$ then $X^T$ now has more columns than rows and can no longer have a kernel of zero. So $X$ is a square matrix coupling $\mu$ and $u^{(2)}$. By the same argument $Y$ is also a $(n-r)\times(n-r)$ matrix, with the first $r$ elements of $v$ now contained in the vector $v^{(1)}$ thanks to an orthogonal transformation of $v$. The single bulk vector fields $u$ and $v$ have each been split into two vectors, with $u$ and $v$ arranged so that
\begin{align}
u =& \left(\begin{matrix}
u^{(1)} \\
u^{(2)}
\end{matrix}\right)
&
v =& \left(\begin{matrix}
v^{(1)} \\
v^{(2)}
\end{matrix}\right)
.\end{align}
The length $r$ vectors $u^{(1)}$ and $v^{(1)}$ do not couple to any of the auxiliary fields and the length $n-r$ vectors $u^{(2)}$ and $v^{(2)}$ couple to the $(n-r)$ auxiliary fields which have not been removed by field redefinitions and do not couple to any other auxiliary fields. We relabel the vector field $\mu$ as $\mu^{(2)}$ to emphasise that it is coupling to the bulk fields in vectors $u^{(2)}$ and $v^{(2)}$ only. So after these field redefinitions the term $\mu_{i,t}X_{ij}u_{j,t}+\mu_{i,t}Y_{ij}v_{j,t}$ has become $\mu^{(2)}_{i,t}X_{ij}u^{(2)}_{j,t}+\mu^{(2)}_{i,t}Y_{ij}v^{(2)}_{j,t}$ with $X$ and $Y$ square with zero kernel. Because they are square with zero kernel both $X$ and $Y$ are invertible, and we can use the field redefinition $\mu^{(2)}\rightarrow \left(X^{-1}\right)^{T}\mu^{(2)}$ to set $X=I$. The condition $XX^T=YY^T$ becomes $YY^T=I$, and so $Y$ must be orthogonal. We no longer have complete freedom to carry out orthogonal transformations on bulk field vectors $u$ and $v$, but orthogonal transformations which do not mix the components of $u^{(1)}$, $v^{(1)}$ with $u^{(2)}$, $v^{(2)}$ are still allowed. So we can use the orthogonal field redefinition $v^{(2)}_i\rightarrow -Y_{ij}^T v^{(2)}_j$ to set $Y=-I$. Finally to keep the type II couplings in the form seen in eq.\eqref{eq:type2} we make the field redefinition $\mu^{(2)}\rightarrow 2\mu^{(2)}$, setting $X=2I$ and $Y=-2I$.

This splitting of the field vectors $u$ and $v$ into $u^{(1)}$ and $u^{(2)}$ and $v^{(1)}$ and $v^{(2)}$ respectively will also require the coupling matrices $A$, $B$ and $C$ to be split up. We take
\begin{align}
A=& \left(\begin{matrix}
A^{(11)} & A^{(12)} \\
-A^{(12)T} & A^{(22)}
\end{matrix}\right)
&
B=& \left(\begin{matrix}
B^{(11)} & B^{(12)} \\
-B^{(12)T} & B^{(22)}
\end{matrix}\right)
&
C=& \left(\begin{matrix}
C^{(11)} & C^{(12)} \\
C^{(21)} & C^{(22)}
\end{matrix}\right)
\end{align}
where $A^{(11)}$, $A^{(22)}$, $B^{(11)}$ and $B^{(22)}$ are antisymmetric to ensure $A$ and $B$ are antisymmetric matrices. The matrices $\tau$, $\phi$, $\pi$ and $\chi$ introduced in eq.\eqref{eq:Pt2} split into
\begin{align}
\tau =& \left(\begin{matrix} \tau^{(1)} & \tau^{(2)} \end{matrix}\right)
&
\phi =& \left(\begin{matrix} \phi^{(1)} & \phi^{(2)} \end{matrix}\right)
&
\pi =& \left(\begin{matrix} \pi^{(1)} & \pi^{(2)} \end{matrix}\right)
&
\chi =& \left(\begin{matrix} \chi^{(1)} & \chi^{(2)} \end{matrix}\right)
.\end{align}

The field redefinition $\mu^{(2)}_{i} \rightarrow \frac{1}{2}\left(C^{(12)T}\right)_{ij}u^{(1)}_{j} +\frac{1}{4}A^{(22)}_{ij}u^{(2)}_{j} +\frac{1}{2}C^{(21)}_{ij}v^{(1)}_{j} -\frac{1}{4}B^{(22)}_{ij}v^{(2)}_{j} +\mu^{(2)}_{i}$ can be used to set $C^{(12)}=A^{(22)}=B^{(22)}=0$. With this simplification the defect Lagrangian can now be written 
\begin{align}
\mathcal{L}^D =& \frac{1}{2}u^{(1)}_{i}A^{(11)}_{ij}u^{(1)}_{j,t} +u^{(1)}_{i}A^{(12)}_{ij}u^{(2)}_{j,t} +\frac{1}{2}v^{(1)}_{i}B^{(11)}_{ij}v^{(1)}_{j,t} +v^{(1)}_{i}B^{(12)}_{ij}v^{(2)}_{j,t} \nonumber \\
&+u^{(1)}_{i}C^{(11)}_{ij}v^{(1)}_{j,t} +u^{(2)}_{i}C^{(22)}_{ij}v^{(2)}_{j,t} +2\mu^{(2)}_{i}\left(u^{(2)}_{i,t} -v^{(2)}_{i,t}\right) +\frac{1}{2}\xi_{i}W_{ij}\xi_{j,t} -F \label{eq:Lredef4}.
\end{align}

Having set the term $\mu_{i,t}\left(XX^T-YY^T\right)_{ij}\mu_{j,t}$ to zero, let us return to the other terms on the right hand side of eq.\eqref{eq:Pt2} which must be a total time derivative for the defect to conserve momentum. The eq.\eqref{eq:Pt2} can now be rewritten as
\begin{align}
\frac{\mathrm{d}P}{\mathrm{d}t} =&
\frac{1}{2}u^{(1)}_{i,t}\left(I-A^{(11)2}-C^{(11)}C^{(11)T}+A^{(12)}A^{(12)T}\right)_{ij}u^{(1)}_{j,t} \nonumber \\
& +\frac{1}{2}u^{(2)}_{i,t}\left(I-C^{(22)}C^{(22)T}+A^{(12)T}A^{(12)}+4\pi^{(2)}\right)_{ij}u^{(2)}_{j,t} \nonumber \\
& -\frac{1}{2}v^{(1)}_{i,t}\left(I-B^{(11)2}-C^{(11)T}C^{(11)}+B^{(12)}B^{(12)T}\right)_{ij}v^{(1)}_{j,t} \nonumber \\
& -\frac{1}{2}v^{(2)}_{i,t}\left(I-C^{(22)T}C^{(22)}+B^{(12)^T}B^{(12)}+4\chi^{(2)}\right)_{ij}v^{(2)}_{j,t} \nonumber \\
& -u^{(1)}_{i,t}\left(A^{(11)}A^{(12)}-2\pi^{(1)T}\right)_{ij}u^{(2)}_{j,t}
+v^{(1)}_{i,t}\left(B^{(11)}A^{(12)}-2\chi^{(1)T}\right)_{ij}v^{(2)}_{j,t} \nonumber \\
& -u^{(1)}_{i,t}\left(A^{(11)}C^{(11)}-C^{(11)}B^{(11)}\right)_{ij}v^{(1)}_{j,t}
+2u^{(2)}_{i,t}\left(\chi^{(2)}-\pi^{(2)T}\right)_{ij}v^{(2)}_{j,t} \nonumber \\
&-u^{(1)}_{i,t}\left(A^{(12)}C^{(22)}-C^{(11)}B^{(12)}+2\pi^{(1)T}\right)_{ij}v^{(2)}_{j,t} \nonumber \\
& +u^{(2)}_{i,t}\left(A^{(12)^T}C^{(11)}-C^{(22)}B^{(12)T}+2\chi^{(1)}\right)_{ij}v^{(1)}_{j,t} \nonumber \\
& +2u^{(1)}_{i,t}A^{(12)}_{ij}\mu^{(2)}_{j,t}
+u^{(1)}_{i,t}\left(\tau^{(1)}W\right)_{ij}\xi_{j,t}
+2v^{(1)}_{i,t}B^{(12)}_{ij}\mu^{(2)}_{j,t}
+v^{(1)}_{i,t}\left(\phi^{(1)T}W\right)_{ij}\xi_{j,t} \nonumber \\
& +2u^{(2)}_{i,t}\left(C^{(22)}+\psi\right)_{ij}\mu^{(2)}_{j,t}
+u^{(2)}_{i,t}\left(2\omega+\tau^{(2)T}W\right)_{ij}\xi_{j,t} \nonumber \\
& -2v^{(2)}_{i,t}\left(C^{(22)T}+\psi\right)_{ij}\mu^{(2)}_{j,t}
-v^{(2)}_{i,t}\left(2\omega-\phi^{(2)T}W\right)_{ij}\xi_{j,t} \nonumber \\
& +u^{(1)}_{i,t}\left(A^{(11)}_{ij}F_{u^{(1)}_j}+A^{(12)}_{ij}F_{u^{(2)}_j}-C^{(11)}_{ij}F_{v^{(1)}_j}-\pi^{(1)T}_{ij}F_{\mu^{(2)}_j}-\tau^{(1)T}_{ij}F_{\xi_j}\right) \nonumber \\
& -u^{(2)}_{i,t}\left(A^{(12)T}_{ij}F_{u^{(1)}_j}+C^{(22)}_{ij}F_{v^{(2)}_j}+\pi^{(2)T}_{ij}F_{\mu^{(2)}_j}+\tau^{(2)T}_{ij}F_{\xi_j}-2\sigma_i\right) \nonumber \\
& -v^{(1)}_{i,t}\left(C^{(11)T}_{ij}F_{u^{(1)}_j}+B^{(11)}_{ij}F_{v^{(1)}_j}+B^{(12)}_{ij}F_{v^{(2)}_j}+\chi^{(1)T}_{ij}F_{\mu^{(2)}_j}+\phi^{(1)T}_{ij}F_{\xi_j}\right) \nonumber \\
& -v^{(2)}_{i,t}\left(C^{(22)T}_{ij}F_{u^{(2)}_j}-B^{(12)T}_{ij}F_{v^{(1)}_j}+\chi^{(2)T}_{ij}F_{\mu^{(2)}_j}+\phi^{(2)T}_{ij}F_{\xi_j}+2\sigma_i\right) \nonumber \\
& +\mu^{(2)}_{i,t}\left(2F_{u^{(2)}_i}+2F_{v^{(2)}_i}-\psi^T_{ij}F_{\mu^{(2)}_j}\right)
-\xi_{i,t}\left(\omega^T_{ij}F_{\mu^{(2)}_j}+W_{ij}\rho_j\right) \nonumber \\
& +\frac{1}{2}\left(\! F_{u^{(1)}_i}F_{u^{(1)}_i}\!+F_{u^{(2)}_i}F_{u^{(2)}_i}\!-F_{v^{(1)}_i}F_{v^{(1)}_i}\!-F_{v^{(2)}_i}F_{v^{(2)}_i}\!\right)
\!- F_{\mu^{(2)}_i}\sigma_i\!- F_{\xi_i}\rho_i\!-\! U\!+\! V \label{eq:Pt4}
.\end{align}
Terms in eq.\eqref{eq:Pt4} containing two $t$ derivatives must be set to zero, as they cannot be written as a total time derivative. From the coefficients of $u^{(1)}_{i,t}\mu^{(2)}_{j,t}$ and $v^{(1)}_{i,t}\mu^{(2)}_{j,t}$ in eq.\eqref{eq:Pt4} we have $A^{(12)}=0$ and $B^{(12)}=0$. The $u^{(1)}_{i,t}\xi_{j,t}$ and $v^{(1)}_{i,t}\xi_{j,t}$ terms set $\tau^{(1)}=0$ and $\phi^{(1)}=0$. The coefficients of $u^{(2)}_{i,t}\xi_{j,t}$ and $v^{(2)}_{i,t}\xi_{j,t}$  constrain $\omega=\frac{1}{2}\phi^{(2)T}W$ and $\tau^{(2)}=-\phi^{(2)}$, whilst we can see that  $\pi^{(1)}=0$ and $\chi^{(1)}=0$ by looking at
the coefficients of $u^{(1)}_{i,t}u^{(2)}_{j,t}$, $v^{(1)}_{i,t}v^{(2)}_{j,t}$, $u^{(1)}_{i,t}v^{(2)}_{j,t}$ and $u^{(2)}_{i,t}v^{(1)}_{j,t}$. For the coefficient of $u^{(2)}_{i,t}v^{(2)}_{j,t}$ to vanish we need that $\chi^{(2)}=\pi^{(2)T}$ and from the coefficients of $u^{(2)}_{i,t}\mu^{(2)}_{j,t}$ and $v^{(2)}_{i,t}\mu^{(2)}_{j,t}$ we find that $\psi=-C^{(22)}$ and that $C^{(22)}$ is symmetric. The field redefinition $\mu_i\rightarrow S_{ij}u^{(2)}_j+S'_{ij}v^{(2)}_j+\mu_i$, where $S$ and $S'$ are symmetric can always be used to set the symmetric part of $C^{(22)}$ (the symmetry of $S$ and $S'$ ensure we do not introduce new terms proportional to $u^{(2)}_{i,t}u^{(2)}_{j,t}$ or $v^{(2)}_{i,t}v^{(2)}_{j,t}$ into the Lagrangian in eq.\eqref{eq:Lredef4}). Since $C^{(22)}$ must be entirely symmetric we can choose to set $C^{(22)}=I$. The vanishing of the coefficients of $u^{(2)}_{i,t}u^{(2)}_{j,t}$ and $v^{(2)}_{i,t}v^{(2)}_{j,t}$ then set $\chi^{(2)}$ and $\pi^{(2)}$ to be antisymmetric. The coefficient of $u^{(1)}_{i,t}u^{(1)}_{j,t}$would be zero if $I-A^{(11)2}-C^{(11)}C^{(11)T}$ could be made antisymmetric, but as it is explicitly symmetric we must set it to zero. Following the method in \cite{bcz04b} we set $C^{(11)}C^{(11)T}=\left(I-A^{(11)}\right)\left(I-A^{(11)T}\right)$. The matrix $A^{(11)}$ is antisymmetric and so has purely imaginary eigenvalues, therefore the matrix $(I-A^{(11)})$ has no zero eigenvalues and we can write $\left(I-A^{(11)}\right)^{-1}C^{(11)}\left(\left(I-A^{(11)}\right)^{-1}C^{(11)}\right)^T=I$. Therefore $\left(I-A^{(11)}\right)^{-1}C^{(11)}=Q$, where $Q$ is an orthogonal matrix and we can set $C^{(11)}=\left(I-A^{(11)}\right)Q$. As previously mentioned we still have the freedom to carry out an orthogonal transformation on $u^{(1)}$ or $v^{(1)}$ without changing the form of the Lagrangian in eq.\eqref{eq:Lredef4}, and we can use such transformations to set $C^{(11)}=\left(I-A^{(11)}\right)$. The condition from the coefficient of  $u^{(1)}_{i,t}v^{(1)}_{j,t}$ is now $A^{(11)}\left(I-A^{(11)}\right)=\left(I-A^{(11)}\right)B^{(11)}$, and as $\left(I-A^{(11)}\right)$ is both invertible and commutes with $A^{(11)}$ we have $B^{(11)}=A^{(11)}$. This also ensures that the coefficient of $v^{(1)}_{i,t}v^{(1)}_{j,t}$ also vanishes. We will set $A^{(11)}=A$ as the superscript is no longer necessary to identify this matrix. All the coupling matrices apart from $A$ have now been set, either to ensure momentum conservation or via field redefinitions.

Putting this all together we have found that in order for a defect to be momentum conserving its Lagrangian must, up to orthogonal transformations of the bulk fields $u$ and $v$ and field redefinitions of the auxiliary fields $\mu$ and $\xi$, be of the form
\begin{align}
\mathcal{L}^D =& \frac{1}{2}u^{(1)}_{i}A_{ij}u^{(1)}_{j,t} +\frac{1}{2}v^{(1)}_{i}A_{ij}v^{(1)}_{j,t} +u^{(1)}_{i}\left(I-A\right)_{ij}v^{(1)}_{j,t} \nonumber \\
&+u^{(2)}_{i}v^{(2)}_{i,t} +2\mu^{(2)}_{i}\left(u^{(2)}_{i,t} -v^{(2)}_{i,t}\right) +\frac{1}{2}\xi_{i}W_{ij}\xi_{j,t} -F \label{eq:Lgeneral}
\end{align}
where $A$ may be any antisymmetric matrix, $W$ is given in eq.\eqref{eq:W} and the components of the bulk vector fields may be divided in any way between the vector fields $u^{(1)}$, $v^{(1)}$ and $u^{(2)}$, $v^{(2)}$. The Lagrangian appears to have split into a type I defect, a type II defect and some extra degrees of freedom, with these separate systems only interacting through the defect potential. Note that if there are no auxiliary fields, so that $\mu^{(2)}$, $\xi$, $u^{(2)}$ and $v^{(2)}$ are absent, then this Lagrangian reduces to the form of the $A_n$ ATFT Toda defect in eq.\eqref{eq:type1}. On the other hand, in the case of a single auxiliary field coupling to single component bulk fields, the fields $u^{(1)}$, $v^{(1)}$ and $\xi$ vanish and the Lagrangian is in the same form as the Lagrangian of the Tzitz\'{e}ica defect \eqref{eq:type2}.

That the defect Lagrangian is in the form eq.\eqref{eq:Lgeneral} is a necessary but not yet a sufficient condition for the defect to be momentum-conserving. So far we have eliminated all the terms in eq.\eqref{eq:Pt4} which are 
quadratic in time derivatives. To ensure that the defect is momentum conserving we must consider the terms which are 
linear or independent of time derivatives; in this way we shall find additional constraints, in particular on the form of 
the defect potential $F$.
Applying the constraints on the coupling matrices which we have just found the momentum conservation condition for the defect becomes
\begin{align}
\frac{\mathrm{d}P}{\mathrm{d}t} =&
u^{(1)}_{i,t}\left(A_{ij}F_{u^{(1)}_j}-\left(I-A\right)_{ij}F_{v^{(1)}_j}\right)
-u^{(2)}_{i,t}\left(F_{v^{(2)}_i}-\pi^{(2)}_{ij}F_{\mu^{(2)}_j}-\phi^{(2)T}_{ij}F_{\xi_j}-2\sigma_i\right) \nonumber \\
& -v^{(1)}_{i,t}\left(\left(I+A\right)_{ij}F_{u^{(1)}_j}+A_{ij}F_{v^{(1)}_j}\right)
-v^{(2)}_{i,t}\left(F_{u^{(2)}_i}+\pi^{(2)}_{ij}F_{\mu^{(2)}_j}+\phi^{(2)T}_{ij}F_{\xi_j}+2\sigma_i\right) \nonumber \\
& +\mu^{(2)}_{i,t}\left(2F_{u^{(2)}_i}+2F_{v^{(2)}_i}+F_{\mu^{(2)}_i}\right)
+\xi_{i,t}\left(\frac{1}{2}W_{ij}\phi^{(2)}_{jk}F_{\mu^{(2)}_k}-W_{ij}\rho_j\right) \nonumber \\
& +\frac{1}{2}\left(F_{u^{(1)}_i}F_{u^{(1)}_i}+F_{u^{(2)}_i}F_{u^{(2)}_i}-F_{v^{(1)}_i}F_{v^{(1)}_i}-F_{v^{(2)}_i}F_{v^{(2)}_i}-2F_{\mu^{(2)}_i}\sigma_i-2F_{\xi_i}\rho_i\right)-U+V \label{eq:Pt5}
.\end{align}
From eq.\eqref{eq:PtOt} we see that the terms involving one $t$ derivative will set the derivatives of the unknown quantity $\Omega$. The terms containing no $t$ derivatives cannot be written as a total time derivative, so must be set to zero. The conditions for momentum conservation are therefore
\begin{align}
\Omega_{u^{(1)}_i} =& -A_{ij}F_{u^{(1)}_j}+\left(I-A\right)_{ij}F_{v^{(1)}_j} \label{eq:mcc1} \\
\Omega_{v^{(1)}_i} =& \left(I+A\right)_{ij}F_{u^{(1)}_j}+A_{ij}F_{v^{(1)}_j} \label{eq:mcc2} \\
\Omega_{u^{(2)}_i} =& F_{v^{(2)}_i}-\pi^{(2)}_{ij}F_{\mu^{(2)}_j}-\phi^{(2)T}_{ij}F_{\xi_j}-2\sigma_i \label{eq:mcc3} \\
\Omega_{v^{(2)}_i} =& F_{u^{(2)}_i}+\pi^{(2)}_{ij}F_{\mu^{(2)}_j}+\phi^{(2)T}_{ij}F_{\xi_j}+2\sigma_i \label{eq:mcc4} \\
\Omega_{\mu^{(2)}_i} =& -2F_{u^{(2)}_i}-2F_{v^{(2)}_i}-F_{\mu^{(2)}_i} \label{eq:mcc5} \\
\Omega_{\xi_{i}} =& -\frac{1}{2}W_{ij}\phi^{(2)}_{jk}F_{\mu^{(2)}_k}+W_{ij}\rho_j \label{eq:mcc6} \\
2(U-V) =& F_{u^{(1)}_i}F_{u^{(1)}_i}+F_{u^{(2)}_i}F_{u^{(2)}_i}-F_{v^{(1)}_i}F_{v^{(1)}_i}-F_{v^{(2)}_i}F_{v^{(2)}_i}-2F_{\mu^{(2)}_i}\sigma_i-2F_{\xi_i}\rho_i \label{eq:mcc7}
\end{align}
where $P+\Omega$ is the conserved momentum-like quantity. At this point we can simplify these momentum conservation conditions significantly by introducing new fields $p=\frac{1}{2}\left(u+v\right)$, $q=\frac{1}{2}\left(u-v\right)$ and new quantities $D$ and $\bar{D}$ with $F=D+\bar{D}$ and $\Omega=D-\bar{D}$. The field vectors $p$ and $q$ split into $p^{(1)}$, $p^{(2)}$ and $q^{(1)}$, $q^{(2)}$ in exactly the same way as the $u$ and $v$ vector fields split into $u^{(1)}$, $u^{(2)}$ and $v^{(1)}$, $v^{(2)}$. The momentum conservation conditions in eqs.\eqref{eq:mcc1}-\eqref{eq:mcc6} then simplify to
\begin{align}
\bar{D}_{p^{(1)}_i} =& 0 \label{eq:Dcond1} \\
\bar{D}_{p^{(2)}_i} =& 0 \label{eq:Dcond2} \\
D_{q^{(1)}_i} =& -A_{ij}D_{p^{(1)}_j} \label{eq:Dcond3} \\
D_{\mu^{(2)}_i} =& -D_{p^{(2)}_i} \label{eq:Dcond4} \\
2\sigma_i =& -D_{q^{(2)}_i}-\pi^{(2)}_{ij}\left(D_{\mu^{(2)}_j}+\bar{D}_{\mu^{(2)}_j}\right)-\phi^{(2)T}_{ij}\left(D_{\xi_j}+\bar{D}_{\xi_j}\right) \label{eq:Dcond5} \\
2\rho_i =& \phi^{(2)}_{ij}\left(D_{\mu^{(2)}_j}+\bar{D}_{\mu^{(2)}_j}\right)-2W_{ij}\left(D_{\xi_j}-\bar{D}_{\xi_j}\right) \label{eq:Dcond6}
.\end{align}
The first four of these equations are satisfied if we require the dependencies of $D$ and $\bar{D}$ to be
\begin{align}
D =& D\left(p^{(1)}+Aq^{(1)},p^{(2)}-\mu^{(2)},q^{(2)},\xi\right) \label{eq:D} \\
\bar{D} =& \bar{D}\left(q^{(1)},q^{(2)},\mu^{(2)},\xi\right) \label{eq:barD}
.\end{align}
The second two equations simply set the two arbitrary vectors $\sigma$ and $\rho$ we introduced previously. Rewriting eq.\eqref{eq:mcc7} using eqs.\eqref{eq:Dcond1}-\eqref{eq:Dcond6} and recalling $A$ and $\pi^{(2)}$ are antisymmetric gives
\begin{align}
2(U-V) =& D_{p^{(1)}_i}\bar{D}_{q^{(1)}_i} +D_{q^{(2)}_i}\bar{D}_{\mu^{(2)}_i} -D_{\mu^{(2)}_i}\bar{D}_{q^{(2)}_i} -4D_{\xi_{i}}W_{ij}\bar{D}_{\xi_{j}} \label{eq:mcc}.
\end{align}
So a momentum conserving defect has a Lagrangian density which can, using field redefinitions, be written in the form given in eq.\eqref{eq:Lgeneral} and a defect potential given by $F=D+\bar{D}$ where quantities $D\left(p^{(1)}+Aq^{(1)},p^{(2)}-\mu^{(2)},q^{(2)},\xi\right)$, $\bar{D}\left(q^{(1)},q^{(2)},\mu^{(2)},\xi\right)$ satisfy the momentum conservation condition in eq.\eqref{eq:mcc}. 
The total conserved energy and momentum of the system are $E+D+\bar{D}$ and $P+D-\bar{D}$, where $E$ and $P$ are the bulk energy and momentum.

A redefinition
$
\mu^{(2)}_i\rightarrow\mu^{(2)}_i+f\left(q^{(2)}\right)_{q^{(2)}_i}
$
does not alter the defect Lagrangian in eq.\eqref{eq:Lgeneral} as it only introduces a total time derivative. Therefore it does not affect the defect equations or any of the subsequent working to find the momentum conservation condition in eq.\eqref{eq:mcc}, and so once $D$ and $\bar{D}$ satisfying the condition have been found these field redefinitions can be used to give other $D$ and $\bar{D}$ satisfying the same momentum conservation condition.

\section{Defects and B\"{a}cklund transformations}

The link between defects and B\"{a}cklund transformations is not surprising, as a B\"{a}cklund transformation is a set of first order equations which, when satisfied, imply that the fields involved also satisfy some other non-linear equations and in constructing the defect equations we are attempting to find some system of first order equations which are satisfied by the same fields as satisfy the non-linear bulk equations of motion.

In \cite{bcz04a,bcz04b} it was noticed that if the defect equations of motion were taken to hold everywhere then they were a B\"{a}cklund transformation for the bulk equations of motion. However, the defect equations for a type II defect do not give a B\"{a}cklund transformation directly. In \cite{cz09b} a new B\"{a}cklund transformation of the Tzitz\'{e}ica model was found by considering the B\"{a}cklund transformation for the type I $A_2$ defect and then folding this model to the Tzitz\'{e}ica model. In doing so the defect equations for a momentum conserving Tzitz\'{e}ica defect are retrieved and an additional equation also appeared. It was noticed that this additional equation was the same as that obtained by taking the momentum conserving defect equations but with $x\leftrightarrow t$ and $\bar{D}\rightarrow-\bar{D}$. Taking the set of defect equations and adding to that the set of defect equations with $x\leftrightarrow t$ and $\bar{D}\rightarrow-\bar{D}$, whilst taking these equations to hold everywhere, gave a B\"{a}cklund transformation for the Tzitz\'{e}ica theory. As we are attempting to find B\"{a}cklund transformations for a general field theory with the bulk Lagrangians as given in eqs.\eqref{eq:bulkLu},\eqref{eq:bulkLv},  which is obviously not obtained by folding $A_n$, this observation is crucial. Note that this procedure applied to type I defect equations leaves them unchanged. The main stumbling block in getting a B\"{a}cklund transformation directly from the type II defect equations is that these equations involve the auxiliary fields, which are only defined at $x=0$. However the procedure described above will introduce $x$ derivatives of these fields to the equations.

The momentum conserving defect Lagrangian is given in eq.\eqref{eq:Pt5}, with $F=D+\bar{D}$ where $D$ and $\bar{D}$ must satisfy eqs.\eqref{eq:D},\eqref{eq:barD},\eqref{eq:mcc}. Using this in the Euler-Lagrange equations gives the defect equations of motion, which we choose to write here in terms of the fields $p=\frac{1}{2}(u+v)$ and $q=\frac{1}{2}(u-v)$ and light cone coordinates $x_{\pm}=\frac{1}{2}(t\pm x)$. We denote $\partial_{x_{\pm}}$ as $\partial_{\pm}$.
\begin{align}
p^{(1)}_{i,-}+A_{ij}q^{(1)}_{j,-} =& \frac{1}{2}\bar{D}_{q^{(1)}_i} \\
p^{(2)}_{i,-}-\mu^{(2)}_{i,+}-\mu^{(2)}_{i,-} =& -\frac{1}{2}\left(D_{q^{(2)}_i}+\bar{D}_{q^{(2)}_i}\right) \\
q^{(1)}_{i,+} =& -\frac{1}{2}D_{p^{(1)}_i} \\
q^{(2)}_{i,+} =& -\frac{1}{2}D_{p^{(2)}_i} \\
q^{(2)}_{i,-} =& \frac{1}{2}\bar{D}_{\mu^{(2)}_i} \\
\xi_{i,+}+\xi_{i,-} =& -2W_{ij}\left(D_{\xi_j}+\bar{D}_{\xi_j}\right).
\end{align}
Carrying out the transformations $x\leftrightarrow t$ (so $\partial_-\rightarrow-\partial_-$) and $\bar{D}\rightarrow-\bar{D}$ gives the additional set of equations
\begin{align}
p^{(1)}_{i,-}+A_{ij}q^{(1)}_{j,-} =& \frac{1}{2}\bar{D}_{q^{(1)}_i} \\
p^{(2)}_{i,-}+\mu^{(2)}_{i,+}-\mu^{(2)}_{i,-} =& \frac{1}{2}\left(D_{q^{(2)}_i}-\bar{D}_{q^{(2)}_i}\right) \\
q^{(1)}_{i,+} =& -\frac{1}{2}D_{p^{(1)}_i} \\
q^{(2)}_{i,+} =& -\frac{1}{2}D_{p^{(2)}_i} \\
q^{(2)}_{i,-} =& \frac{1}{2}\bar{D}_{\mu^{(2)}_i} \\
\xi_{i,+}-\xi_{i,-} =& -2W_{ij}\left(D_{\xi_j}-\bar{D}_{\xi_j}\right).
\end{align}
Taking both sets of equations to hold simultaneously and over all space rather than just at $x=0$ we can remove any repeated equations. Rearranging the remaining equations to simplify them gives
\begin{align}
p^{(1)}_{i,-}+A_{ij}q^{(1)}_{j,-} =& \frac{1}{2}\bar{D}_{q^{(1)}_i} \label{eq:Bt1} \\
p^{(2)}_{i,-}-\mu^{(2)}_{i,-} =& -\frac{1}{2}\bar{D}_{q^{(2)}_i} \label{eq:Bt2} \\
q^{(1)}_{i,+} =& -\frac{1}{2}D_{p^{(1)}_i} \label{eq:Bt3} \\
q^{(2)}_{i,+} =& -\frac{1}{2}D_{p^{(2)}_i} \label{eq:Bt4} \\
q^{(2)}_{i,-} =& \frac{1}{2}\bar{D}_{\mu^{(2)}_i} \label{eq:Bt5} \\
\mu^{(2)}_{i,+} =& \frac{1}{2}D_{q^{(2)}_i} \label{eq:Bt6} \\
\xi_{i,+} =& -2W_{ij}D_{\xi_j} \label{eq:Bt7} \\
\xi_{i,-} =& -2W_{ij}\bar{D}_{\xi_j} \label{eq:Bt8}.
\end{align}
Cross-differentiating these equations and using the dependencies of $D$ and $\bar{D}$ given in eqs.\eqref{eq:D},\eqref{eq:barD} and the fact that $D$ and $\bar{D}$ must obey the momentum conservation condition in eq.\eqref{eq:mcc} we can easily see that these give the bulk equations of motion for field vectors $p$ and $q$, plus some bulk equations of motion for what were the auxiliary fields.

So the systems of equations $u_{itt}-u_{ixx}+U(u)=0$ and $v_{itt}-v_{ixx}+V(v)=0$ where $u=p+q$, $v=p-q$ have a B\"{a}cklund transformation given by eqs.\eqref{eq:Bt1}-\eqref{eq:Bt8} if quantities $D\left(p^{(1)}+Aq^{(1)},p^{(2)}-\mu^{(2)},q^{(2)},\xi\right)$ and $\bar{D}\left(q^{(1)},q^{(2)},\mu^{(2)},\xi\right)$ can be found which satisfy eq.\eqref{eq:mcc}. Here $A$ can be any antisymmetric matrix, $W$ is given by eq.\eqref{eq:W}, the bulk fields may be divided between $p^{(1)}$, $q^{(1)}$ and $p^{(2)}$, $q^{(2)}$ in any way and the auxiliary fields may be divided between $\mu^{(2)}$ and $\xi$ in any way as long as $p^{(1)}$ and $q^{(1)}$ are the same length, $p^{(2)}$, $q^{(2)}$ and $\mu^{(2)}$ are the same length and $\xi$ contains an even number of fields due to the form of the matrix $W$.

\section{Defects in affine Toda field theories}

An ATFT is described by the Lagrangian density \begin{align}
\mathcal{L}_u = \frac{1}{2}u_{i,t}u_{i,t} -\frac{1}{2}u_{i,x}u_{i,x} -U
\quad\quad\quad\quad
U = \frac{m^2}{\beta^2} \sum_{i=0}^r n_i e^{\beta(\alpha_i)_ju_j} \label{eq:Lbulk}
\end{align}
where $\alpha_i$, $i=1,...,n$ are the simple root vectors of a Lie algebra, $n_i$, $i=1,...,n$ are a set of integers characteristic of each algebra, $n_0=1$ and $\alpha_0=-\sum_{i=1}^r n_i \alpha_i$ gives the root which corresponds to the extra node on an affine Dynkin diagram \cite{mop81,ot83}. $m$ is the mass constant, $\beta$ is the coupling constant and as they are unimportant in the classical case we set $m=\beta=1$. The vector $u=(u_1,...,u_n)^T$ lies in the space spanned by the simple root vectors and the fields $\{u_i\}$ are the projections of $u$ onto the basis of this vector space. Recall that the components of the vector $u$ appear in the vector $u^{(1)}$ if they do not couple to the auxiliary field $\mu^{(2)}$ and in the vector $u^{(2)}$ if they do couple to $\mu^{(2)}$. Call the vector space in which $u^{(1)}$ (and $v^{(1)}$) live the 1-space and the vector space in which $u^{(2)}$ (and $v^{(2)}$ and $\mu^{(2)}$) live the 2-space. The vector $u^{(1)}$ can be thought of as a projection of $u$ onto the 1-space and $u^{(2)}$ as the projection of $u$ onto the 2-space. The 1- and 2-space are orthogonal and sum to the vector space in which the vector $u$ lives, that is, the space spanned by the simple root vectors. Therefore we can have $\alpha^{(1)}_i$ as the projection of a simple root $\alpha_i$ onto the 1-space and $\alpha^{(2)}_i$ as the projection onto the 2-space.

All ATFTs have been shown to be integrable \cite{mik79,mop81,ot85} and soliton solutions 
have been found for all ATFTs \cite{mcg94b}
. In order for soliton solutions to exist the potential $U$ must have multiple vacua, and so the exponent within the potential must take imaginary values as $x\rightarrow\pm\infty$ (specifically it must be $2\pi n i$ so we can use the definition of $\alpha_0$ to ensure $U_{u_i}=0$). Normally the constant $\beta$ would be taken to be purely imaginary, but as we have set $\beta=1$ we instead allow the field $u$ to be complex, taking an appropriate purely imaginary value as $x\rightarrow\pm\infty$.

Defects were first introduced in \cite{bcz04a}, where a momentum conserving type I sine-Gordon defect was found. This type I defect was generalised to allow any number of bulk fields in \cite{bcz04b}, a Lax pair was constructed for the system with a defect, and it was found that requiring the defect to be momentum conserving gave the same constraints as requiring the defect to be integrable. In the cases of solitons and integrable boundaries the results for ATFTs based on different algebras were fairly similar. However, in \cite{bcz04b} it was shown that the momentum conservation condition found for they type I defects could only be satisfied if the bulk ATFTs were based on the $A_n$ series. In \cite{cz09b} the introduction of an extra degree of freedom at the defect to give a type II defect admitted a momentum conserving Tzitz\'{e}ica defect and here we hope that the generalisation of the type II defect to one with any number of bulk and auxiliary fields will allow momentum conserving defects for all ATFTs. 

When considering the general defect found in section \ref{sec:momconsv} with a particular potential the fact that we carried out rotations on the bulk fields in order to simplify the defect Lagrangian becomes relevant. Fortunately, for the potential given in eq.\eqref{eq:Lbulk} the fact that the root vectors $\alpha_i$ are defined only by their inner products with each other means that rotations of $u$ and $v$ do not fundamentally change $U$ and $V$. Take $\{\alpha_i\}$ to be the simple root vectors fixed to be certain, reasonably simple vectors. Over the course of the previous section the bulk fields have undergone the transformations $u\rightarrow Qu$ and $v\rightarrow Q'v$, where $Q$ and $Q'$ are some orthogonal transformations which we do not specify here. But the sets of simple roots $\{Q^T\alpha_i\}$ and $\{Q'^T\alpha_i\}$ have the same Dynkin diagram as $\{\alpha_i\}$, and so we can begin with $U$ dependent on $\{Q^T\alpha_i\}$ and $V$ dependent on $\{Q'^T\alpha_i\}$. After $u$ and $v$ have undergone their field redefinitions both $U$ and $V$ will be dependent on $\{\alpha_i\}$, but will still be the bulk potentials for the same ATFT we started with.

By considering the exponentials of the field $p$ in the momentum conservation condition in eq.\eqref{eq:mcc} when we use the potentials $U$ and $V$ as given in eq.\eqref{eq:Lbulk}, and the dependencies of $D$ and $\bar{D}$ in eqs.\eqref{eq:D},\eqref{eq:barD}, we see that they must take the form
\begin{align}
D =& \sigma\sum_{i=0}^n x_i\left(q^{(2)},\xi\right) e^{ \left(\alpha_i\right)^{(1)}_{j}\left(p^{(1)}_{j}+A_{jk}q^{(1)}_{k}\right) +(\alpha_i)^{(2)}_{j}\left(p^{(2)}_{j}-\mu^{(2)}_j\right)} \label{eq:Dgeneral} \\
\bar{D} =& \frac{1}{\sigma}\sum_{i=0}^n y_i\left(q^{(1)},q^{(2)},\xi\right) e^{ -(\alpha_i)^{(1)}_{j} A_{jk}q^{(1)}_{k} +(\alpha_i)^{(2)}_{j}\mu^{(2)}_j } \label{eq:Dbargeneral}
\end{align}
where $\sigma$ is a constant and $x_i$ and $y_i$ are functions yet to be determined.

There is no obvious systematic way of ensuring that $D$ and $\bar{D}$ satisfy the momentum conservation condition in eq.\eqref{eq:mcc} for a particular set of simple roots. In particular, we have not yet found any systematic way of splitting the root space into the 1- and 2-spaces. Instead we have used trial and error to find momentum conserving defects for some ATFTs.

\subsection{\boldmath$D_4$ defect} \label{sec:D4}

For an ATFT based on the root vectors of $D_4$ we choose to use
\begin{align}
\alpha_0 =& \left(\begin{matrix}
-1 \\
-1 \\
0 \\
0 \end{matrix}\right)
&
\alpha_1 =& \left(\begin{matrix}
1 \\
-1 \\
0 \\
0 \end{matrix}\right)
&
\alpha_2 =& \left(\begin{matrix}
0 \\
1 \\
-1 \\
0 \end{matrix}\right)
&
\alpha_3 =& \left(\begin{matrix}
0 \\
0 \\
1 \\
-1 \end{matrix}\right)
&
\alpha_4 =& \left(\begin{matrix}
0 \\
0 \\
1 \\
1 \end{matrix}\right)
.\end{align}
The bulk potentials are then \cite{mop81}
\begin{align}
U =& e^{-u_1-u_2} +e^{u_1-u_2} +2e^{u_2-u_3} +e^{u_3-u_4} +e^{u_3+u_4} \label{eq:D4U} \\
V =& e^{-v_1-v_2} +e^{v_1-v_2} +2e^{v_2-v_3} +e^{v_3-v_4} +e^{v_3+v_4} \label{eq:D4V}
.\end{align}

Through trial and error it was found that when $A=0$, there are no $\xi$ fields and the basis of the 1-space is $(e_1,e_4)$ (so $u^{(1)}=\left( u_1 \: u_4 \right)^T$) and the basis of the 2-space is $(e_2,e_3)$ (so $u^{(2)}=\left( u_2 \: u_3 \right)^T$) the Lagrangian in eq.\eqref{eq:Lgeneral} gives a momentum conserving defect. Written out explicitly the momentum conserving defect Lagrangian for $D_4$ ATFT is
\begin{align}
\mathcal{L}^D =& u_1v_{1,t} +u_2v_{2,t} +u_3v_{3,t} +u_4v_{4,t} +2\mu_2(u_{2,t}-v_{2,t}) +2\mu_3(u_{3,t}-v_{3,t}) -F \label{eq:LD4}
.\end{align}
Recalling that $p_i=\frac{1}{2}(u_i+v_i)$, $q_i=\frac{1}{2}(u_i-v_i)$ the momentum conservation condition in eq.\eqref{eq:mcc} in this case is
\begin{align}
2(U-V) =& D_{p_1}\bar{D}_{q_1} +D_{q_2}\bar{D}_{\mu_2} -D_{\mu_2}\bar{D}_{q_2} +D_{q_3}\bar{D}_{\mu_3} -D_{\mu_3}\bar{D}_{q_3} +D_{p_4}\bar{D}_{q_4} \label{eq:D4momconsv}
\end{align}
and is satisfied by
\begin{align}
D =& \sigma \Big( \left(e^{p_1}+e^{-p_1}\right)\left(e^{q_2}+e^{-q_2}\right)e^{-p_2+\mu_2} +2\left(e^{q_3}+e^{-q_3}\right)e^{p_2-p_3-\mu_2+\mu_3} \nonumber \\
& \quad\quad +\left(e^{p_4}+e^{-p_4}\right)e^{p_3-\mu_3} \Big) \label{eq:D4D} \\
\bar{D} =& \frac{1}{\sigma} \Big( \left(e^{q_1}+e^{-q_1}\right)e^{-\mu_2} +\left(e^{q_2}+e^{-q_2}\right)e^{\mu_2-\mu_3} +\left(e^{q_3}+e^{-q_3}\right)\left(e^{q_4}+e^{-q_4}\right)e^{\mu_3} \Big) \label{eq:D4Dbar}
.\end{align}
A field redefinition of $\mu_2\rightarrow\mu_2+f(q_2,q_3)_{q_2}$ and $\mu_3\rightarrow\mu_3+f(q_2,q_3)_{q_3}$ would not change the form of the Lagrangian in eq.\eqref{eq:LD4}, and so would not affect any of the working up to the momentum conservation condition. It would only affect the form of $D$ and $\bar{D}$. Therefore we can use this redefinition to give a family of defect potentials which all satisfy the same momentum conservation condition.

This momentum conserving defect can be used to give a B\"{a}cklund transformation for the $D_4$ ATFT as discussed in the previous section. 

Using the method introduced in \cite{cz09b} and expanded on in \cite{rob14a}, whereby a procedure involving squeezing together several defects and then folding the associated Dynkin diagram gives rise to a defect in the folded ATFT, it should be possible to use this $D_4$ defect to construct a $G_2$ defect, however this has not yet been achieved.

\subsection{\boldmath$D_4$ solitons and defects}

As in \cite{mcg94b} the solitons are given by
\begin{align}
u=-\sum_{i=0}^n \alpha_i\ln{\tau_i}
\end{align}
where the $\tau$ functions are dependent on $E=e^{\sqrt{\lambda}(\cosh{\theta}x-\sinh{\theta}t)+c}$ with $\lambda$ and $c$ being constants. For $D_4$ there is one soliton with $\lambda=2$ which is associated with the central node on the Dynkin diagram and three with $\lambda=6$ which are associated with the outer nodes on the Dynkin diagram. The $\tau$ functions of these solitons are
\begin{align}
\lambda=2 : \qquad \tau_0&=\tau_1=1+E \quad \tau_3=\tau_4=1-E \quad \tau_2=1+E^2 \label{eq:soliton21} \\
\tau_0&=\tau_3=1+E \quad \tau_1=\tau_4=1-E \quad \tau_2=1+E^2 \label{eq:soliton23} \\
\tau_0&=\tau_4=1+E \quad \tau_1=\tau_3=1-E \quad \tau_2=1+E^2 \label{eq:soliton24} \\
\lambda=6 : \qquad \tau_0&=\tau_1=\tau_3=\tau_4=1+E \quad \tau_2=1-4E+E^2 \label{eq:soliton6} 
.
\end{align}

Solitons appear when the field interpolates between two vacua in the potential, given in eq.\eqref{eq:Lbulk} for ATFTs. Here a vacuum occurs when the vector field is a weight of $D_4$ multiplied by $2\pi ni$. Weight lattices are associated with roots, and depending on which lattice the soliton vacuum is in we say it is associated with that root. For $D_4$ it is possible to change which root a soliton with $\lambda=2$ is associated with by an orthogonal transformation of the field.

In both the quantum and linearised classical cases the type I defects were found to be purely transmitting \cite{dms94b,kl99,bcz04a}, so here we are considering purely transmitting defects and expect the soliton to be delayed by the defect. We do not consider whether a soliton associated with one root can be transformed by the defect to a soliton associated with a different root. To find the delays from the defect we write the soliton to the right of the defect in terms of $E$ and the soliton to the left in terms of $zE$, where $z$ is the delay. The defect equations can then be solved for $z$ to give the delay experienced by the soliton as it passes through the defect.

For the soliton given in eq.\eqref{eq:soliton21} the possible delays are
\begin{align}
&z=\frac{(1-\rho)(\omega-\rho)}{(1+\rho)(\omega+\rho)} & &\text{or} &\frac{(1-\rho)(\omega^2-\rho)}{(1+\rho)(\omega^2+\rho)} & &\text{or} & &\frac{(\omega-\rho)(\omega^2-\rho)}{(\omega+\rho)(\omega^2+\rho)}
\end{align}
with $\rho=2^{\frac{1}{6}}\sigma e^{\theta}$ and $\omega=e^{\frac{2\pi i}{3}}=\frac{1}{2}(-1+\sqrt{3}i)$, so all powers of $\omega$ are cube roots of unity. $\rho$ and $\omega$ take these values in all the soliton delays calculated here. For the soliton in eq.\eqref{eq:soliton23} the delays are
\begin{align}
&z=\frac{1+\rho}{1-\rho} & &\text{or} &\frac{\omega+\rho}{\omega-\rho} & &\text{or} & &\frac{\omega^2+\rho}{\omega^2-\rho}.
\end{align}
The delays are identical for the soliton in eq.\eqref{eq:soliton24}. Finally for the soliton in eq.\eqref{eq:soliton6} the possible delays are
\begin{align}
&z=\frac{(i-\rho)(i\omega+\rho)}{(i+\rho)(i\omega-\rho)} & &\text{or} &\frac{(i+\rho)(i\omega^2-\rho)}{(i-\rho)(i\omega^2+\rho)} & &\text{or} & &\frac{(i\omega-\rho)(i\omega^2+\rho)}{(i\omega+\rho)(i\omega^2-\rho)}.
\end{align}
Note that for all the sets of delays taking $\rho\rightarrow \omega\rho$ cycles through the possible delays.

The values of $\rho$ which correspond to a pole or a zero in the soliton delay give the defect parameter and soliton rapidity which lead to the soliton being absorbed by the defect. The same phenomenon was observed for sine-Gordon solitons passing through a defect in \cite{bcz04a}. The absorption of a soliton by the defect makes sense in terms of B\"{a}cklund transformations, as B\"{a}cklund transformations can be used to obtain a $n$ soliton solution from an $n-1$ soliton solution.

\subsection{\boldmath$D_n$ defect} \label{sec:Dn}

For the $D_n$ ATFT the potential is given by \cite{mop81}
\begin{align}
U =& e^{-u_1-u_2} +e^{u_1-u_2} +\sum_{i=2}^{n-2} 2e^{u_i-u_{i+1}} +e^{u_{n-1}-u_{n-1}} +e^{u_{n-1}+u_n} \label{eq:DnU} \\
V =& e^{-v_1-v_2} +e^{v_1-v_2} +\sum_{i=2}^{n-2} 2e^{v_i-v_{i+1}} +e^{v_{n-1}-v_{n-1}} +e^{v_{n-1}+v_n} \label{eq:DnV}
.\end{align}
In the $D_4$ defect the fields which appeared in the type II part of the defect were those which appeared in the term in the potential associated with the central node on the Dynkin diagram. To move from $D_4$ to $D_n$ we assume that the fields appearing in the terms associated with the central chain of nodes on the Dynkin diagram will appear in the type II part of the defect. So we take $A=0$, no $\xi$ fields and that the basis of the 1-space is $(e_1,e_n)$ and the basis of the 2-space is $(e_2 ... e_{n-1})$. The momentum conserving defect Lagrangian is given by inserting these values into eq.\eqref{eq:Lgeneral} and the momentum conservation condition in eq.\eqref{eq:mcc} is satisfied by
\begin{align}
D =& \sigma\Bigg(\!\left(e^{p_1}+e^{-p_1}\right)\left(e^{q_2}+e^{-q_2}\right)e^{-p_2+\mu_2} +2\sum_{i=2}^{n-2} \left(e^{q_{i+1}}+e^{-q_{i+1}}\right)e^{p_i-p_{i+1}-\mu_i+\mu_{i+1}} \nonumber \\
&\quad\; +\left(e^{p_n}+e^{-p_n}\right)e^{p_{n-1}-\mu_{n-1}} \Bigg) \label{eq:DnD} \\
\bar{D} =& \frac{1}{\sigma} \Bigg(\! \left(e^{q_1}+e^{-q_1}\right)e^{-\mu_2} \sum_{i=2}^{n-2} \left(e^{q_i}+e^{-q_i}\right)e^{\mu_i-\mu_{i+1}} \nonumber \\
&\quad\; +\left(e^{q_{n-1}}+e^{-q_{n-1}}\right)\left(e^{q_n}+e^{-q_n}\right)e^{\mu_{n-1}} \Bigg) \label{eq:DnDbar}
.\end{align}
As at the end of subsection \ref{sec:D4} redefinitions of the $\mu_i$ fields can be used to give different defect potentials satisfying the same momentum conservation condition.

\subsection{\boldmath$A_n$ defect} \label{sec:An}

The potential of the ATFT based on $A_n$ may be written as \cite{mop81} 
\begin{align}
U =& e^{-u_1+u_{n+1}} +\sum_{i=1}^{n} e^{u_i-u_{i+1}} \label{eq:AnU} \\
V =& e^{-v_1+v_{n+1}} +\sum_{i=1}^{n} e^{v_i-v_{i+1}} \label{eq:AnV}
.\end{align}
The fields in the $A_n$ ATFT have the additional constraint $\sum_{i=1}^{n+1} u_i =0$, $\sum_{i=1}^{n+1} v_i =0$. This potential is entirely made up of terms similar to those associated with the central chain of nodes in $D_n$, and so we take $A=0$, no $\xi$ fields and the 2-space covers the whole vector space spanned by the simple roots. That is, there are the same number of auxiliary fields as there are bulk fields, with the same condition on them. The momentum conservation condition in eq.\eqref{eq:mcc} is satisfied by
\begin{align}
D =& \sigma\left( \sum_{i=1}^{n} \left(e^{q_{i+1}}+e^{-q_{i+1}}\right)e^{p_i-p_{i+1}-\mu_i+\mu_{i+1}} +\left(e^{q_1}+e^{-q_1}\right)e^{p_{n+1}-p_1-\mu_{n+1}+\mu_1} \right) \label{eq:AnD} \\
\bar{D} =&\frac{1}{\sigma}\left( \sum_{i=1}^{n}\left(e^{q_i}+e^{-q_i}\right)e^{\mu_i-\mu_{i+1}} +\left(e^{q_{n+1}}+e^{-q_{n+1}}\right)e^{\mu_{n+1}-\mu_1} \right) \label{eq:AnDbar}
.\end{align}
This is the same as the defect given by squeezing two $A_n$ defects together \cite{cz09b, rob14a}.

\subsection{\boldmath$B_n$ defect} \label{sec:Bn}

The potential of the ATFT based on $B_n$ may be written as \cite{mop81}
\begin{align}
U =& e^{-u_1-u_2} +e^{u_1-u_2} +\sum_{i=2}^{n-1} 2e^{u_i-u_{i+1}} +2e^{u_n} \label{eq:BnU}
.\end{align}
Taking $A=0$, no $\xi$ fields and the basis of the 1-space to be $(e_1)$ and the basis of the 2-space to be $(e_2 ... e_n)$ gives a momentum conserving defect. The momentum conservation condition in eq.\eqref{eq:mcc} is satisfied by
\begin{align}
D =& \sigma\Bigg(\!\!\left(e^{p_1}+e^{-p_1}\right)\left(e^{q_2}+e^{-q_2}\right)e^{-p_2+\lambda_2} \nonumber \\
&\quad\;+2\sum_{i=1}^{n} \left(e^{q_{i+1}}+e^{-q_{i+1}}\right)e^{p_i-p_{i+1}-\lambda_i+\lambda_{i+1}} +2e^{p_n-\lambda_n} \Bigg) \label{eq:BnD} \\
\bar{D} =& \frac{1}{\sigma}\left(\!\left(e^{q_1}+e^{-q_1}\right)e^{-\lambda_2} +\sum_{i=1}^{n}\left(e^{q_i}+e^{-q_i}\right)e^{\lambda_i-\lambda_{i+1}} +\left(e^{q_n}+e^{-q_n}\right)e^{\lambda_n} \right) \label{eq:BnDbar}
.\end{align}

\subsection{\boldmath$C_n$ defect} \label{sec:Cn}

The bulk potential of the ATFT baased on $C_n$ may be written as \cite{mop81}
\begin{align}
U =& e^{-2 u_1} +\sum_{i=1}^{n-1} 2e^{u_i-u_{i+1}} +e^{2 u_n}
\label{eq:BnU2}
.\end{align}
For a momentum conserving defect we take $A=0$, no $\xi$ fields and that the 2-space covers the whole vector space spanned by the simple roots. The momentum conservation condition in eq.\eqref{eq:mcc} is satisfied by
\begin{align}
D =& \sigma\Bigg(\! \left(e^{q_1}+e^{-q_1}\right)^2e^{-2p_1+2\mu_1} \nonumber \\
&\quad\;+2\sum_{i=1}^{n-1} \left(e^{q_{i+1}}+e^{-q_{i+1}}\right)e^{p_i-p_{i+1}-\mu_i+\mu_{i+1}} +e^{2p_n-2\mu_n} \Bigg) \label{eq:CnD} \\
\bar{D} =& \frac{1}{\sigma}\left(e^{-2\mu_1} +\sum_{i=1}^{n-1}\left(e^{q_i}+e^{-q_i}\right)e^{\mu_i-\mu_{i+1}} +\left(e^{q_n}+e^{-q_n}\right)^2e^{2\mu_n} \right) \label{eq:CnDbar}
.\end{align}
For $C_2$ this momentum conserving defect is the same as that found in \cite{rob14a} by squeezing together $A_n$ type I defects and then carrying out a folding procedure.

\section{Conclusions} \label{sec:conclusions}

This work has confirmed previous results (the squeezed sine-Gordon defects found in \cite{cz09b} and the $C_3$ defects found in \cite{rob14a}), provided new energy and momentum conserving defects, and gives us a framework which will hopefully cover all defects in ATFTs. The fact that all defects satisfying the conditions given in this paper can be used to give a B\"{a}cklund transformation suggests that these momentum conserving defects are also integrable, as well as being interesting in its own right. The explicit calculations for transmission of solitons through the $D_4$ defect also strongly suggest that it is an integrable system.

The obvious next step is to attempt to find defects in the remaining exceptional simply laced ATFTs ($E_6$, $E_7$, $E_8$). In principle these are the only remaining cases it is necessary to solve, as all non simply laced ATFTs can be found by folding simply laced ATFTs and the folding procedure for defects in \cite{rob14a} can then be used to find momentum conserving defects, and so B\"{a}cklund transformations, for all ATFTs. These momentum conserving defects have not been found so far due to the difficultly of finding appropriate 2-space. It may also be that a non-zero $A$ matrix or $\xi$ vector field is required. However we have no systematic way of finding the 1- and 2-space splitting, $A$ matrix or $\xi$ field required for a momentum conserving defect and this is a difficult task to complete by trial and error alone.

The existence of a Lax pair for the defects found so far would confirm the integrability of a system with a defect. Studying Lax pairs has the added advantage of potentially giving us some more insight into the structure of these defects in ATFTs, and work currently being carried out on this problem by one of the authors \cite{bri17} has already yielded some pointers towards the correct 1- and 2-space splitting for the $E$ series defects. Similar work has already been carried out for a boundary and a defect in the nonlinear Schr\"{o}dinger model \cite{zam14} as well as for the type I defects in ATFTs in \cite{bcz04b,cz09a,hk08,doi15}.

The $D_4$ defect is the simplest of the new defects found, and so is the obvious candidate for investigating how these defects behave when interacting with solitons. Although we have checked that solitons are transmitted by the defect we yet to make investigations into the details about the topological charge of the defect before and after a soliton has passed through it, whether the defect can change  one soliton into another soliton with different topological charge, and the behaviour of the auxiliary fields during soliton transmission. We intend to pursue this further.

Such investigations into the interactions between classical solitons and defects are likely to be necessary for our final suggested angle of continuation, finding the transmission matrix for a quantum defect. Quantum multi-field defects were investigated in \cite{cz07}, a single quantum auxiliary field was investigated in \cite{cz10}, and quantum type II $A_n$ defects and quantum algebras relating to other ATFTs were investigated in \cite{cz11}. Finding the quantum transmission matrix for the defects in this paper would probably involve combining these ideas.

\bibliographystyle{JHEP}
\bibliography{allreferences}

\end{document}